\documentclass[review]{elsarticle}
\usepackage{geometry}
\newcommand*{\ARXIV}{}

\usepackage[acronym,nonumberlist]{glossaries} 

\usepackage{amsmath,amssymb,amscd,latexsym,dsfont,wasysym,bbm}
\usepackage{natbib}  
\biboptions{numbers,sort&compress}
\usepackage{amsmath} 
\usepackage[caption=false,font=footnotesize]{subfig} 
\usepackage[normalem]{ulem} 

\usepackage{enumitem}
\newlist{Aenumerate}{enumerate}{1}
\setlist[Aenumerate]{label=A.\arabic*}

\usepackage{booktabs}

\usepackage{microtype}

\usepackage{hyperref}

\usepackage{comment}

\newcommand{\CFilesBib}{Common.Files.Bib}
\usepackage{amsmath,amssymb,amscd,latexsym,dsfont,wasysym,bbm}
\usepackage{float}
\usepackage{color}
\usepackage{multirow,multicol} 
\usepackage{verbatim}
\usepackage{psfrag}
\usepackage{comment}
\usepackage{makeidx}
\usepackage[]{algorithm}
\usepackage{algorithmicx}
\usepackage{algpseudocode}
\usepackage{cases}
 \usepackage{setspace}
\usepackage{pgfplots} 

\newcommand{\tr}[1]{\textrm{#1}}
\newcommand{\mr}[1]{\mathrm{#1}}
\newcommand{\tnr}[1]{{\textnormal{#1}}}

\newcommand{\mc}[1]{\mathcal{#1}}
\newcommand{\mf}[1]{\mathsf{#1}}
 
\newcommand{\ms}[1]{\mathds{#1}}

\newcommand{\ov}[1]{\overline{#1}}

\newcommand{\ba}{\boldsymbol{a}}

\newcommand{\bd}{\boldsymbol{d}}

\newcommand{\be}{\boldsymbol{e}}

\newcommand{\bh}{\boldsymbol{h}}

\newcommand{\bm}{\boldsymbol{m}}

\newcommand{\br}{\boldsymbol{r}}

\newcommand{\bu}{\boldsymbol{u}}

\newcommand{\bv}{\boldsymbol{v}}
\newcommand{\bw}{\boldsymbol{w}}

\newcommand{\bx}{\boldsymbol{x}}

\newcommand{\by}{\boldsymbol{y}}

\newcommand{\bz}{\boldsymbol{z}}

\newcommand{\bzero}{\boldsymbol{0}}

\newcommand{\blambda}{\boldsymbol{\lambda}}

\newcommand{\figref}[1]{Fig.~\ref{#1}}

\newcommand{\ie}{i.e.,~} 		
\newcommand{\eg}{e.g.,~}	

\newcommand{\argmax}{\mathop{\mr{argmax}}}
\newcommand{\argmin}{\mathop{\mr{argmin}}}

\newcommand{\set}[1]{\{#1\}}

\newcommand{\cd}{\cdot}
\newcommand{\ld}{\ldots}

\newcommand{\e}{\mr{e}}

\newcommand{\pdf}{f}            			

\newcommand{\Ex}{\ms{E}}     			
\newcommand{\T}{^{\mf{T}}}            		
\newcommand{\Th}{^{\mr{H}}}             		
\newcommand{\dd}{\,\mr{d}}             		

\newcommand{\mcN}{\mc{N}}

\newcommand{\mfm}{\mf{m}}

\newcommand{\Real}{\mathbb{R}}		
\newcommand{\Complex}{\mathbb{C}}		

\newcommand{\matA}{\tnr{\textbf{A}}}

\newcommand{\matI}{\tnr{\textbf{I}}}

\newcommand{\matQ}{\tnr{\textbf{Q}}}
\newcommand{\matR}{\tnr{\textbf{R}}}
\newcommand{\matS}{\tnr{\textbf{S}}}

\newcommand{\matV}{\tnr{\textbf{V}}}

\newcommand{\matX}{\tnr{\textbf{X}}}

\pgfplotsset{compat=1.12}
\usetikzlibrary{positioning,shadows,shapes,fit,arrows,backgrounds,trees}

\tikzstyle{rect_my} = [draw, rectangle, minimum width=2cm, text width=1.8cm, fill=gray!15, 
  text centered,  minimum height=.9cm]
\tikzstyle{square_my} = [draw, rectangle, minimum width=1cm, text width=0.8cm, fill=gray!15, 
  text centered,  minimum height=.9cm]
\tikzstyle{square_my_graph} = [draw, rectangle, minimum width=1.2cm, text width=1cm, fill=gray!15, 
  text centered,  minimum height=1.2cm]
\tikzstyle{circle_my} = [draw, circle, minimum width=1cm, text width=0.8cm, fill=gray!15, 
  text centered,  minimum height=.9cm]
\tikzstyle{circle_my_graph} = [draw, circle, minimum width=1.1cm, text width=.8cm, fill=gray!15, 
  text centered]
\tikzstyle{cloud_my} = [draw, shape=cloud, minimum width=1cm, text width=0.8cm, fill=gray!15, 
  text centered,  minimum height=1cm]

\tikzstyle{point_my} = [draw=none, minimum width=0cm, text width=0cm, fill=none, 
  text centered,  minimum height=0cm]    
\tikzstyle{line_my} = [draw, -latex]    
\tikzstyle{box_my}=[draw, minimum size=2em, text width=4.5em, text centered]
\tikzstyle{bigbox_my}=[draw, inner sep=15pt]
\tikzstyle{arrow_my} = [thick,->,>=stealth]
\tikzstyle{noarrow_my} = [thick,-,=>stealth]

\usepackage[acronym,nonumberlist]{glossaries} 
\usepackage{hyperref}
\usepackage{empheq}
\usepackage{verbatim}
\usepackage{graphicx}
\newacronym[\glsshortpluralkey=PDFs,\glslongpluralkey=probability density functions]{pdf}{PDF}{probability density function}
\newacronym[\glsshortpluralkey=CDFs,\glslongpluralkey=cumulative density functions]{cdf}{CDF}{cumulative density function}
\newacronym[\glsshortpluralkey=CCDFs,\glslongpluralkey=complementary cumulative density functions]{ccdf}{CDF}{complementary cumulative density function}
\newacronym[\glsshortpluralkey=PMFs,\glslongpluralkey=probability mass functions]{pmf}{PMF}{probability mass function}
\newacronym[]{lhs}{l.h.s.}{left-hand side}
\newacronym[]{rhs}{r.h.s.}{right-hand side} 

\newacronym[]{bicm}{BICM}{bit-interleaved coded modulation}
\newacronym[]{bicmid}{BICM-ID}{BICM with iterative demapping}
\newacronym[]{cm}{CM}{coded modulation}
\newacronym[]{tcm}{TCM}{trellis-coded modulation}
\newacronym[]{mlc}{MLC}{multi-level coding}
\newacronym[]{pam}{PAM}{pulse amplitude modulation}
\newacronym[]{bpsk}{BPSK}{binary phase shift keying}
\newacronym[]{qam}{QAM}{quadrature amplitude modulation}
\newacronym[]{16qam}{16-QAM}{16-points quadrature amplitude modulation}
\newacronym[]{psk}{PSK}{phase shift keying}
\newacronym[\glsshortpluralkey=LLRs,\glslongpluralkey=logarithmic likelihood ratios]{llr}{LLR}{logarithmic likelihood ratio}
\newacronym[]{oc}{OC}{operating characteristic}

\newacronym[\glsshortpluralkey=MIs,\glslongpluralkey=mutual informations]{mi}{MI}{mutual information}
\newacronym[\glsshortpluralkey=GMIs,\glslongpluralkey=generalized mutual informations]{gmi}{GMI}{generalized mutual information}
\newacronym[]{eesm}{EESM}{exponential effective-SNR-mapping}
\newacronym[]{bicm-gmi}{BICM-GMI}{BICM generalized mutual information}
\newacronym[]{awgn}{AWGN}{additive white Gaussian noise}
\newacronym[]{bsc}{BSC}{binary symetric channel}
\newacronym[]{amc}{AMC}{adaptive modulation and coding}
\newacronym[]{csi}{CSI}{channel state information}
\newacronym[]{cqi}{CQI}{channel quality indicator}
\newacronym[]{kl}{KL}{Kullback-Leibler}
\newacronym[]{cmm}{CMM}{circular moment matching}
\newacronym[]{ga}{GA}{Gaussian approximation}

\newacronym[]{sp}{SP}{set-partitioning}
\newacronym[]{gsm}{GSM}{global system for mobile communications}
\newacronym[]{edge}{EDGE}{enhanced data rates for GSM evolution}
\newacronym[]{3gpp}{3GPP}{3rd generation partnership project}
\newacronym[]{umts}{UMTS}{Universal Mobile Telecommunication System}
\newacronym[]{lte}{LTE}{Long Term Evolution}
\newacronym[]{dvb}{DVB}{digital video broadcasting}
\newacronym[]{fdd}{FDD}{Frequency Division Duplexing}

\newacronym[\glsshortpluralkey=CCs,\glslongpluralkey=convolutional codes]{cc}{CC}{convolutional code}
\newacronym[\glsshortpluralkey=PCCCs,\glslongpluralkey=parallel concatenated convolutional codes]{pccc}{PCCC}{parallel concatenated convolutional code}
\newacronym[\glsshortpluralkey=TCs,\glslongpluralkey=turbo codes]{tc}{TC}{turbo code}
\newacronym{ldpc}{LDPC}{low-density parity-check}
\newacronym[]{ofdm}{OFDM}{orthogonal frequency-division multiplexing}
\newacronym[]{bep}{BEP}{bit-error probability}
\newacronym[]{wep}{WEP}{word-error probability}
\newacronym[]{sep}{SEP}{symbol-error probability}
\newacronym[]{pep}{PEP}{pairwise-error probability}
\newacronym[]{ttcm}{TTCM}{turbo-trellis coded modulation}
\newacronym[]{uep}{UEP}{unequal error protection}
\newacronym[\glsshortpluralkey=CENCs,\glslongpluralkey=convolutional encoders]{cenc}{CENC}{convolutional encoder}
\newacronym[]{mimo}{MIMO}{multiple-input multiple-output}
\newacronym[\glsshortpluralkey=SNRs,\glslongpluralkey=signal-to-noise ratios]{snr}{SNR}{signal-to-noise ratio}
\newacronym[\glsshortpluralkey=SINRs,\glslongpluralkey=signal-to-interference-plus-noise ratios]{sinr}{SINR}{signal-to-interference-plus-noise ratio}
\newacronym[]{msb}{MSB}{most-significative bit}
\newacronym[]{bcjr}{BCJR}{Bahl--Cocke--Jelinek--Raviv}
\newacronym[]{cbc}{CBC}{Colavolpe--Barbieri--Caire}
\newacronym[]{skr}{SKR}{Shayovitz--Kreimer--Raphaeli}
\newacronym[\glsshortpluralkey=SEDs,\glslongpluralkey=squared Euclidean distances]{sed}{SED}{squared Euclidean distance}
\newacronym[\glsshortpluralkey=EDs,\glslongpluralkey=Euclidean distances]{ed}{ED}{Euclidean distance}
\newacronym[\glsshortpluralkey=MEDs,\glslongpluralkey=minimum Euclidean distances]{med}{MED}{minimum Euclidean distance}
\newacronym[]{core}{CoRe}{constellation rearrangement}
\newacronym[]{pdl}{PDL}{parallel decoding of the individual levels}
\newacronym[\glsshortpluralkey=GCs,\glslongpluralkey=Gray codes]{gc}{GC}{Gray code}
\newacronym[]{brgc}{BRGC}{binary-reflected Gray code}
\newacronym[]{nbc}{NBC}{natural binary code}
\newacronym[]{fbc}{FBC}{folded-binary code}
\newacronym[]{bsgc}{BSGC}{binary semi-Gray code}
\newacronym[]{msp}{MSP}{modified set-partitioning}
\newacronym[]{ssp}{SSP}{semi set-partitioning}
\newacronym[]{fhd}{FHD}{free Hamming distance}
\newacronym[]{mfhd}{MFHD}{maximum free Hamming distance}
\newacronym[]{ods}{ODS}{optimal distance spectrum}
\newacronym[]{iud}{i.u.d.}{independent and uniformly distributed}
\newacronym[]{ud}{u.d.}{uniformly distributed}
\newacronym[]{iid}{i.i.d.}{independent, identically distributed}
\newacronym[]{ami}{AMI}{accumulated mutual information}
\newacronym[]{bico}{BICO}{binary-input continuous-output}
\newacronym[]{gh}{GH}{Gauss--Hermite}
\newacronym[]{gum}{GUM}{Gaussian--uniform mixture}

\newacronym[\glsshortpluralkey=BSs,\glslongpluralkey=base-stations]{bs}{BS}{base-station}
\newacronym[\glsshortpluralkey=MSs,\glslongpluralkey=mobile-stations]{ms}{MS}{mobile-stations}

\newacronym[]{phy}{PHY}{physical layer} 
\newacronym[]{rlc}{RLC}{Radio-Link control} 
\newacronym[]{ran}{RAN}{Radio Access Network} 
\newacronym[]{llc}{LLC}{logical link control} 
\newacronym[]{tcp}{TCP}{transmission control protocol} 
\newacronym[]{mac}{MAC}{media access control} 
\newacronym[]{fft}{FFT}{fast Fourier transform} 
\newacronym[]{ft}{FT}{Fourrier transform}
\newacronym[]{cf}{CF}{characteristic function} 
\newacronym[]{mgf}{MGF}{moment generating function} 
\newacronym[]{ee}{EE}{energy efficiency} 
\newacronym[]{eb}{EB}{energy per bit}
\newacronym[]{kkt}{KKT}{Karush--Kuhn--Tucker} 
\newacronym[]{mcs}{MCS}{modulation/coding scheme} 
\newacronym[]{fec}{FEC}{forward error correction}
\newacronym[]{arq}{ARQ}{automatic repeat request}
\newacronym[]{harq}{HARQ}{hybrid ARQ}
\newacronym[]{tarq}{TARQ}{truncated HARQ}
\newacronym[]{ir}{IR}{incremental redundancy}
\newacronym[]{rpr}{RR}{repetition redundancy}
\newacronym[]{rrharq}{RR-HARQ}{repetition redundancy HARQ}
\newacronym[]{irharq}{IR-HARQ}{incremental redundancy HARQ}
\newacronym[]{ack}{ACK}{positive acknowledgment}
\newacronym[]{nack}{NACK}{negative acknowledgment}
\newacronym[]{hol}{HoL}{head of the line}
\newacronym[]{crc}{CRC}{cyclic redundancy check}
\newacronym[]{dp}{DP}{dynamic programming}
\newacronym[]{gp}{GP}{geometric programming}
\newacronym[]{per}{PER}{packet error rate}
\newacronym[]{ber}{BER}{bit error rate}
\newacronym[]{op}{OP}{outage probability}
\newacronym[]{spa}{SPA}{saddle-point approximation}
\newacronym[]{mrc}{MRC}{maximum ratio combining}
\newacronym[]{mdp}{MDP}{Markov decision process}
\newacronym[]{lp}{LP}{linear programming}
\newacronym[]{pomdp}{POMDP}{partially observable Markov decision process}
\newacronym[]{psimdp}{PSI-MDP}{partial state information Markov decision process}
\newacronym[]{scpp}{SCPP}{stochastic shortest path problem}

\newacronym[]{forw}{frwd}{forward}
\newacronym[]{feed}{fdbk}{feedback}

\newacronym[]{mm}{MM-HARQ}{multi-message HARQ}
\newacronym[]{xp}{XP-HARQ}{cross-packet HARQ}
\newacronym[]{ts}{TS}{time-sharing}
\newacronym[]{sc}{SC}{superposition coding}
\newacronym[]{sbrq}{SBRQ}{systematic backward retransmission}
\newacronym[]{brq}{BRQ}{backward retransmission}
\newacronym[]{lharq}{L-HARQ}{layer-coded HARQ}
\newacronym[]{anlharq}{AoN-HARQ}{all-or-none L-HARQ}
\newacronym[]{vlharq}{VL-HARQ}{variable-length HARQ}

\newacronym[]{pp}{PPP}{point process}
\newacronym[]{ppp}{PPP}{Poisson point process}

\newacronym[]{fide}{FIDE}{F\'ed\'eration Internationale des \'Echecs}
\newacronym[]{fifa}{FIFA}{F\'ed\'eration Internationale de Football Association}
\newacronym[]{fivb}{FIVB}{F\'ed\'eration Internationale de Volleyball}
\newacronym[]{epl}{EPL}{English Premier League}
\newacronym[]{nhl}{NHL}{National Hockey League}
\newacronym[]{shl}{SHL}{Swedish Hockey League}
\newacronym[]{nfl}{NFL}{National Football League}
\newacronym[]{ipl}{IPL}{Indian Premier League}
\newacronym[]{nba}{NBA}{National Basketball Association}
\newacronym[]{mls}{MLS}{Major League Soccer}

\newacronym[]{sg}{SG}{stochastic gradient}
\newacronym[]{lms}{LMS}{least mean squares}
\newacronym[]{rls}{RLS}{recursive least squares}
\newacronym[]{vss}{VSS}{variable step-size}
\newacronym[]{hfa}{HFA}{home-field advantage}
\newacronym[]{ha}{HA}{home advantage}
\newacronym[]{mov}{MOV}{margin of victory}
\newacronym[]{ac}{AC}{adjacent categories}
\newacronym[]{cl}{CL}{cumulative link}
\newacronym[]{glm}{GLM}{generalized linear models}
\newacronym[]{nn}{NN}{neural networks}
\newacronym[]{rps}{RPS}{ranked probability score}
\newacronym[]{mse}{MSE}{mean square error}
\newacronym[]{mmse}{MMSE}{minimum mean square error}
\newacronym[]{rmse}{RMSE}{root mean squared error}
\newacronym[]{ols}{OLS}{ordinary least squares}
\newacronym[]{map}{MAP}{maximum a posteriori}
\newacronym[]{ml}{ML}{maximum likelihood}
\newacronym[]{loo}{LOO}{leave-one-out}
\newacronym[]{alo}{ALO}{approximate leave-one-out}
\newacronym[]{logo}{LOGO}{leave-one-game-out}
\newacronym[]{alogo}{ALOGO}{approximate leave-one-game-out}
\newacronym[]{msd}{MSD}{mean-square deviation}
\newacronym[]{lop}{LOP}{linear ordering problem}
\newacronym[]{so}{SO}{shootouts}
\newacronym[]{rt}{RT}{regulation time}
\newacronym[]{ot}{OT}{overtime}
\newacronym[]{rr}{RR}{round-robin}
\newacronym[]{irt}{IRT}{item-response theory}

\newacronym[]{dmp}{DMP}{discretized message passing}
\newacronym[]{mp}{MP}{message passing}
\newacronym[]{ep}{EP}{expectation propagation}
\newacronym[]{em}{EM}{expectation maximization}
\newacronym[]{hmm}{HMM}{hiden Markov models}

\newacronym[]{svd}{SVD}{singular values decomposition}

\newacronym[]{skf}{SKF}{simplified Kalman filter}
\newacronym[]{vskf}{vSKF}{\emph{vector-covariance} Simplified Kalman Filter}
\newacronym[]{sskf}{sSKF}{\emph{scalar-covariance} Simplified Kalman Filter}
\newacronym[]{fskf}{fSKF}{\emph{fixed-variance} Simplified Kalman Filter}
\newacronym[]{kf}{KF}{Kalman filter}
\newacronym[]{gelo}{G-Elo}{generalized Elo}
\newacronym[]{mvdr}{MVDR}{minimum variance distortionless response}
\newacronym[]{lcmv}{LCMV}{linearly-constrained minimum variance}
\newacronym[]{music}{MUSIC}{multiple signal classification}
\newacronym[]{cp}{CP}{canonical polyadic}

\newacronym[]{tpb}{TPB}{tensor-product-basis}

\newtheorem{proposition}{Proposition}

\usepackage{xcolor}
\definecolor{cblue}{HTML}{1965B0}

\definecolor{cred}{HTML}{B8221E}

\definecolor{dgreen}{rgb}{0,0.6,0}

\definecolor{dorange}{RGB}{255, 128, 0}

\definecolor{burntorange}{rgb}{0.8, 0.33, 0.0}

\begin{document}

\begin{frontmatter}




\title{Automatic Regularization for Linear MMSE Filters}


\author[1]{Daniel~Gomes~de~Pinho~Zanco}
\ead{daniel.zanco@inrs.ca}

\author[1]{Leszek~Szczecinski}
\ead{leszek.szczecinski@inrs.ca}

\author[1]{Jacob~Benesty}
\ead{jacob.benesty@inrs.ca}

\address[1]{INRS–Institut National de la Recherche  Scientific, Montreal, QC, H5A-1K6, Canada.}

\begin{abstract}
In this work, we consider the problem of regularization in the design of \acrfull{mmse} linear filters. Using the relationship with statistical machine learning methods, using a Bayesian approach, the regularization parameter is found from the observed signals in a simple and automatic manner. The proposed approach is illustrated in system identification and beamforming examples, where the automatic regularization is shown to yield near-optimal results.
\end{abstract}


\end{frontmatter}

\section{Introduction}
\Gls{mmse} linear filters are ubiquitous in many signal processing applications such as channel equalization \cite[Ch.~5.4]{Sayed08_Book}, system identification \cite{Dogariu21}, antenna beamforming \cite[Ch.~6.5]{Sayed08_Book}, and many others. 

The two main classes of \gls{mmse} filters are (i) the error minimization, where the linear filter is designed to approximate the desired signal with the smallest average squared error, and (ii) interference suppression, where the objective is to minimize the interference energy while maintaining the energy of the desired signal.

The equations solved to obtain the \gls{mmse} filters rely on the implicit or explicit inversion of the covariance matrix of the input signal. To avoid numerical problems and to guarantee the uniqueness of the solution, the equations must be \emph{regularized}, as is most often done by adding a positive regularization parameter to the diagonal elements of the covariance matrix.

Determining the regularization parameter is frequently regarded as a challenge for practitioners and, depending on the \gls{snr} or the type of problem, it is often handcrafted for each specific problem. This attitude changes and, recently, the regularization received in-depth attention in the context of system identification \cite{Pillonetto22_book}. 

On the other hand, this issue is rather well known in the contexts of machine learning and regression analysis, where methods such as cross-validation \cite{Allen71,Golub79} or \gls{em} \cite[18.1.3]{Barber12_Book} are often used to find parameters which are not of direct interest, but affect the solutions (known as \emph{hyperparameters}).

However, despite regularization being crucial to finding \gls{mmse} filters, the signal processing literature, in general does not use simple and general solutions from the area of machine learning. The main reason, we believe, is that they are not offered in closed form and, in general, may require searching over the entire space of solutions and solving the regularized equations multiple times. We show that, in practice, the solution can be found very efficiently via fixed-point iteration and does not entail any significant complexity increase if we exploit the eigenvalue decomposition of the covariance matrix.

This paper is organized as follows. We start with the general problem formulation in Sec.~\ref{Sec:Problem.formulation} and, in Sec.~\ref{Sec:find.alpha} we reformulate it using the probabilistic framework, which allows us to apply the \gls{ml} estimation to the parameters defining the model and obtain the optimal regularization parameter. Section~\ref{Sec:Regularization.for.MVDR} discusses automatic regularization in the interference-suppression problem.
In Sec.~\ref{Sec:Numerical.examples}, to illustrate the operation of the proposed method, we apply it to system identification (as an example of error-minimization) and to beamforming (as an example of interference suppression) to show how the automatically regularized \gls{mmse} filters compare to other methods proposed in the literature and to an ``oracle" solution. The latter relies on ex-ante knowledge of the best regularization parameter, and is obtained by grid search over the space of the latter, by maximizing the performance criterion of interest, which is possible in the simulations where we know all signals involved.

The examples indicate that the regularization parameter, which we find, automatically adjusts to the changes in operational conditions (such as the \gls{snr}) and to the problem structure.

Our main conclusion is that, by adopting the machine learning approach, the automatic regularization is so simple that it deserves to be a go-to solution in the signal processing context.

\section{\gls{mmse} problem formulation}\label{Sec:Problem.formulation}

We consider the linear filtering of the input signal $\bx(t) \in \Complex^M$ using the weights/filter $\bw$ aiming at the approximation of the desired signal $d(t) \in \Complex$. There are two categories of this problem with respect to how the filter $\bw$ is found, which are described below.
\begin{itemize}
\item \textbf{The error-minimization problem}, where we know the desired signal $d(t)$, the filtering error is given by
\begin{align}
\label{linear.model}
e(t) &= d(t) - \bw\Th\bx(t), \quad  t=0,1,\ldots,N-1,
\end{align}
and the \gls{mmse} problem consists in solving
\begin{align}
\label{hat.h}
\hat\bw &= \argmin_{\bw} \left\{ \Ex\left[|d(t) - \bw\Th\bx(t)|^2\right]+ \alpha\|\bw\|_2^2 \right\}\\
\label{wiener.linear.system}
&= (\ov\matR_{\bx} + \alpha \matI)^{-1}\ov{\br}_{\bx d},
\end{align}
where $\Ex[\cdot]$ denotes mathematical expectation taken with respect to all random variables, $\alpha \ge 0$ is a regularization parameter, $|| \cdot ||_2$ is the Euclidean norm, $\ov\matR_{\bx} = \Ex\left[\bx(t)\bx\Th(t)\right]$, $\matI$ is the identity matrix, and $\ov{\br}_{\bx d} = \Ex\left[\bx(t) d^*(t)\right]$; we use $(\cd)\Th$ to denote conjugate-transpose operation, and $(\cd)^*$ denotes complex conjugation.

\item \textbf{The interference suppression problem}, where we assume that the signal $\bx(t)$ has the form:
\begin{align}
\label{x.t.MVDR}
    \bx(t) = d(t)\ba +\bz(t)\in\Complex^M,
\end{align}
with $\bz(t)$ being the interference, and $\ba$ the response generated by the desired signal $d(t)$, where $\|\ba\|^2=M$.
The goal is then to minimize the (energy of) interference in the filtered output $\bw\Th\bx(t)$, \ie
\begin{align}
\label{MVDR.problem.definition}
    \hat\bw & = \argmin_{\bw} \left\{ \Ex\left[|\bw\Th\bx(t)|^2\right] +\alpha\|\bw\|^2 \right\}  \quad \tnr{s. \ t.}\quad \bw\Th\ba=1,
\end{align}
while maintaining the energy of the desired signal, as enforced by the constraint $\bw\Th\ba = 1$.
The problem \eqref{MVDR.problem.definition} is known to be solved by \cite[Sec.~2.8]{Haykin02_Book}
\begin{align}
\label{hat.w.MVDR}
\hat\bw & = \frac{\big(\ov\matR_{\bx}+\alpha\matI\big)^{-1}\ba}{\ba\Th\big(\ov\matR_{\bx}+\alpha\matI\big)^{-1}\ba}.
\end{align}
\end{itemize}

Numerous applications of these two problems have been presented in the literature. For example, the error minimization problem \eqref{wiener.linear.system} is found in system identification, equalization \cite[Ch.~2]{Haykin02_Book}, interference cancellation \cite[Ch.~8]{Tse2005}, and many others. The interference suppression problem \eqref{MVDR.problem.definition} is popular in beamforming \cite{Li03} and spectral estimation \cite{Li96}. 

Note that \eqref{wiener.linear.system} is a regularized version of the Wiener equation \cite[Ch.~2.4]{Haykin02_Book} and \eqref{hat.w.MVDR} is the regularized version of the \gls{lcmv} filter \cite[Ch.~2.8]{Haykin02_Book}.
However, in textbook formulations, the problems \eqref{hat.h} or \eqref{MVDR.problem.definition} are defined with $\alpha=0$, \ie without regularization. The latter is added in \eqref{wiener.linear.system} and \eqref{hat.w.MVDR} by practitioners \cite[Ch.~8.10]{Haykin02_Book}, \cite[Sec.~4]{Dogariu21}, \cite[Sec.~2.B]{Du10} with the aim of improving conditioning of the matrix $\ov{\matR}_{\bx} + \alpha \matI$, which must be inverted (at least implicitly\footnote{The explicit inversion of the matrix in \eqref{wiener.linear.system} may be avoided by solving linear equations \mbox{$(\ov\matR_{\bx} + \alpha \matI)\hat\bw = \br_{\bx d}$}.}). 

The main reason why regularization is required comes from the fact that, in practice, we do not have access to $\ov\matR_{\bx}$ or $\ov\br_{\bx d}$. Rather, they are estimated from the data using time-averaging, 
\begin{align}\label{eq:left.Rx}
\ov\matR_{\bx} &\approx \matR_{\bx}=\frac{1}{N} \sum_{t=0}^{N-1}\bx(t)\bx\Th(t),\\
\label{eq:left.rxd}
\ov{\br}_{\bx d} &\approx \br_{\bx d} =\frac{1}{N} \sum_{t=0}^{N-1}\bx(t)d^*(t).
\end{align}

Then, the regularization term, $\alpha\matI$, is a practical solution to deal with imperfect estimates \eqref{eq:left.Rx}-\eqref{eq:left.rxd}, and/or with the numerical errors involved in solving \eqref{wiener.linear.system}. The parameter $\alpha$ has to be ``appropriately chosen" and will depend on all the elements of the model \eqref{linear.model}. In particular, since the importance of the estimation errors in \eqref{eq:left.Rx}-\eqref{eq:left.rxd} decreases with $N$, we expect that the value of $\alpha$ also decreases with $N$.

\subsection{Known regularization solutions in signal processing}

Recognizing regularization to be an important practical element in the definition of linear filters, this problem was addressed in the literature, particularly in the context of the \gls{mvdr} formulation; two, the most representative solutions, are shown below.

\subsubsection{Ledoit-Wolf matrix shrinkage}

The Ledoit-Wolf matrix shrinkage method \cite{Ledoit04} assumes the following relationship between the true and empirical covariance matrix
\begin{align}
\ov\matR_{\bx} &\approx \beta\matR_{\bx} + \eta\matI,
\end{align}
and, by minimizing the squared Frobenius norm of the approximation error:
\begin{align}
\hat\eta, \hat\beta = \min_{\beta, \eta} \Ex[\|\ov\matR_{\bx} - \beta\matR_{\bx} - \eta\matI\|^2_{\mr{F}}],
\end{align}
finds the shrinkage parameters as \cite[Eqs.~(32)-(33)]{Du10}
\begin{align}
\hat\eta &= \min\left[1, \frac{\hat\rho}{\|\matR_{\bx} - \hat\nu \matI\|^2_{\mr{F}}}\right] \hat\nu,\\
\hat\beta &= 1 - \frac{\hat\eta}{\hat\nu},
\end{align}
where
\begin{align}
\hat\rho &= \frac{1}{N^2} \sum_{t=0}^{N-1} \|\bx(t)\|^4 - \frac{1}{N} \|\matR_{\bx}\|^2_{\mr{F}},\\
\hat\nu &= \frac{1}{M} \tr{Tr}(\matR_{\bx}),
\end{align}
with $\tr{Tr}(\cdot)$ being the trace of a square matrix.

By factorizing $\beta$, the shrinkage parameters can then be converted back into a regularization parameter:
\begin{align}
\alpha_{\tr{LW}} = \frac{\eta}{\beta}.
\end{align}

This method has been used to find regularization in the interference suppression problem \eqref{hat.w.MVDR}, \eg in \cite{Du10}. On the other hand, we are not aware of its application to the error-minimization problem \eqref{wiener.linear.system}, most likely because the latter depends not only on the noisy covariance matrix $\matR_{\bx}$ but, also, explicitly requires a noisy cross-correlation vector $\br_{\bx d}$. 

In that regard, the interference suppression problem uses noisy $\matR_{\bx}$ and error-free $\ba$, and, therefore, appears to be affected only by errors in the former. As we will see, such an interpretation is misleading, and the regularization depends also on $\ba$.\footnote{Note that, in some works, \eg \cite{Li03}, the problem is formulated assuming that $\ba$ is also corrupted by errors. We do not use such a model, as assuming that $\ba$ is perfectly known allows us to emphasize the fact that the regularization depends not only on the noise but also on the deterministic elements of the model.}

\subsubsection{Hoerl, Kennard, and Baldwin regularization}\label{Sec:HKB}

Some regularization strategies are derived by exploiting the fact that the Wiener equations \eqref{wiener.linear.system} can be obtained from the regularized \gls{ols} problem:
\begin{align}
\label{OLS.regularized}
    \hat\bw(\alpha) = \argmin_{\bw} \left[ \frac{1}{N}\| \bd - \matX\Th\bw\|^2 + \alpha\|\bw\|^2 \right],
\end{align}
where $\bd=[d^*(0),\ld,d^*(N-1)]\T$ and $\matX=[\bx(0),\ld,\bx(N-1)]\T$; $(\cd)\T$ is a transposition operator.


The method proposed by Hoerl, Kennard, and Baldwin (HKB) in \cite[Eq.~(2.2)]{ArthurE.Hoerl1975}, finds the regularization in two steps. First, \eqref{OLS.regularized} is solved for $\alpha=0$ and, next, the regularization parameter is calculated as
\begin{align}
\label{alpha.HKB}
\alpha_{\tr{HKB}} &= \frac{\tilde\sigma_e^2(0)}{N \tilde\sigma_w^2(0)},
\end{align}
where
\begin{align}
\label{hat.sigma.e.HKL}
\tilde\sigma_e^2(\alpha) &= \frac{1}{N}\| \bd - \matX\Th\hat\bw(\alpha)\|^2,\\
\label{hat.sigma.w.HKL}
\tilde\sigma_w^2(\alpha) & = \frac{1}{\gamma}\|\hat\bw(\alpha)\|^2,
\end{align}
and $\gamma\in(0,M]$ is the number of degrees of freedom of the solution. In the error-minimization problem, we set $\gamma=M$, while in the interference suppression problem, due to a linear constraint on $\bw$, we set $\gamma=M-1$.


The HKB regularization was studied in the beamforming context \cite{Du10}, but we immediately see that it cannot be applied for $N < M$. This is because then the rank of $\matX$ is smaller than $M$, so there is an infinite number of $\hat\bw(0)$ that solve \eqref{OLS.regularized}, each of which yields $\tilde\sigma_e^2(0) = 0$. Thus, for $N < M$, \eqref{alpha.HKB} produces $\alpha_{\tr{HKB}} = 0$. 

\section{Bayesian formulation and inference of regularization parameter}\label{Sec:find.alpha}


To obtain the Bayesian formulation of the problem, we rewrite \eqref{linear.model} in vector form:
\begin{align}
\label{d.X.h.z}
\bd &=\matX\Th\bw +\be,
\end{align}
where $\bd$, $\matX$ are already defined in \eqref{OLS.regularized}, and $\be=[e^*(0),\ld,e^*(N-1)]\T$.

Assuming that $e(t)$ are \gls{iid} zero-mean Gaussian variables with variance $v_{e}$, we have
\begin{align}
\label{pdf.d|X,h}
\pdf(\bd|\matX,\bw) &= \mcN(\bd; \matX\Th\bw,v_e\matI),
\end{align}
where 
\begin{align}
\label{Gaussian.pdf}
\mcN(\by;\bm,\matV) = \frac{1}{\tr{det}(\pi\matV)} \exp[-(\by-\bm)\Th\matV^{-1}(\by-\bm)]
\end{align}
denotes the \gls{pdf} of a circular, complex Gaussian with mean $\bm$ and covariance matrix $\matV$.

The Bayesian approach models the parameter $\bw$ as a random vector with posterior distribution given by
\begin{align}
\label{posterior.bh}
\pdf(\bw|\bd, \matX) & \propto 
\pdf(\bd|\matX,\bw)\pdf(\bw).
\end{align}

Then, assuming the elements $\bw$ to be \gls{iid} zero-mean, Gaussian random variables with variance $v_w$, \ie
\begin{align}
\label{pdf.h}
\pdf(\bw)
&= \mcN(\bw;\bzero,v_w\matI),
\end{align}
it is simple to see that, using \eqref{pdf.d|X,h} and \eqref{pdf.h}, the posterior distribution \eqref{posterior.bh} is given by
\begin{align}
\pdf(\bw|\bd,\matX) &= \mcN(\bw;\hat\bw,\matR_{\bw}),
\end{align}
where
\begin{align}
\label{mat.K}    
\matR_{\bw}
&=\frac{v_{e}}{N}(\matR_{\bx} + \alpha \matI)^{-1},\\
\label{hat.bw.posterior}
\hat\bw 
&= (\matR_{\bx}+\alpha\matI)^{-1}\br_{\bx d},\\
\label{alpha=ve/vw}
\alpha &= \frac{v_{e}}{Nv_w}.
\end{align}

Of course, \eqref{hat.bw.posterior} being the mean of the posterior, it is also the \gls{map} estimate, \ie $\hat\bw = \argmax_{\bw}\pdf(\bw|\bd,\matX)$ and is the same as the solution of the Wiener equation \eqref{hat.h} obtained from empirical moments given in \eqref{eq:left.Rx}-\eqref{eq:left.rxd}.

This modeling approach is well-known in signal processing textbooks. For example,
\cite[Ch.~4]{Pillonetto22_book} or \cite[Part~VII~-~Summary~and~Notes]{Sayed08_Book} note the equivalence between the \gls{map} estimation of $\bw$ and the Wiener (least-squares) solution. On the other hand, the signal processing literature does not exploit this model to its full extent and does not find the parameters $\bv=[v_w,v_{e}]$ even if it would give us the immediate advantage of defining the regularization parameter $\alpha$ via \eqref{alpha=ve/vw}. An additional advantage is that, knowing $\bv$, we can find the posterior variance $\matR_{\bw}$ which allows us to assess the uncertainty of the estimation: remember, the diagonal elements of $\matR_{\bw}$ are the posterior variances of the estimates $\hat\bw$. 




\subsection{Inference}\label{Sec:Inference}
We will infer the parameters $\bv$ using the \gls{ml} approach:
\begin{align}
\label{hat.bv}
\alpha_{\tnr{ML}}, \hat{v}_e  &=\argmax_{\alpha,v_e} L(\alpha,v_e),\\
L(\alpha,v_e) & = -\log\pdf(\bd|\matX,\bv),
\end{align}
where, instead of $v_w$ and $v_e$, we parameterized the variables using $\alpha=v_e/(N v_w)$, which does not affect the optimality of \gls{ml} solution, and focuses directly on the regularization parameter $\alpha$ we are interested in.\footnote{Of course, we can obtain the \gls{ml} estimates $\hat{v}_{e}$ and $\hat{v}_{w}$, too.} 

We marginalize over $\bw$ to obtain
\begin{align}
\pdf(\bd|\matX,\bv)
&=
\int \pdf(\bd|\matX,\bw,\bv)\pdf(\bw|\bv)\dd \bw,
\end{align}
with the distributions under integration being  those shown in \eqref{pdf.d|X,h} and \eqref{pdf.h}; the conditioning on $\bv$ merely makes explicit their dependence on the parameters $\bv$. Since all the variables are Gaussian, it is rather easy to show that
\begin{align}
\pdf(\bd|\matX,\bv) 
&= 
\frac{\tr{det}[(\matR_{\bx} + \alpha\matI)^{-1}]{\alpha^M}}
{\pi^{N} v_e^N} 
\exp\left[ \frac{N}{v_e} \big(\br_{\bx d}\Th\hat\bw - \tilde\sigma_d^2\big) \right],
\end{align}
where $\tilde\sigma_d^2 = \|\bd\|^2/N$ is the estimate of the second moment of $d(t)$, and, from \eqref{wiener.linear.system}, $\br_{\bx d}\Th\hat\bw$ is real.

Thus,
\begin{equation}
\begin{aligned}
L(\alpha, v_e) 
=& -\log\tr{det}\left[(\matR_{\bx} + \alpha \matI)^{-1}\right] - M\log\alpha + N\log v_e \\
&+ \frac{N}{v_e} (\tilde\sigma_d^2 - \br_{\bx d}\Th\hat\bw) + N\log\pi,
\end{aligned}
\end{equation}
which, for a given $\alpha$, is uniquely minimized by $\hat{v}_e$ satisfying
\begin{align}
\frac{\dd}{\dd v_e} L(\alpha, \hat{v}_e) &= \frac{N}{\hat{v}_e}\left[1 - \frac{1}{\hat{v}_e} (\tilde\sigma_d^2 - \br_{\bx d}\Th\hat\bw)\right] = 0,\\
\hat{v}_e &= \tilde\sigma_d^2 - \br_{\bx d}\Th\hat\bw.
\end{align}

Then, \eqref{hat.bv} is reduced to 
\begin{align}
\label{alpha.ML}
\alpha_{\tnr{ML}} & = \argmin_\alpha L(\alpha),\\
L(\alpha) &= L(\alpha, \hat{v}_e)\\
\label{L(alpha)}
&= -\log\tr{det}\left[(\matR_{\bx} + \alpha \matI)^{-1}\right] - M\log\alpha + N\log(\tilde\sigma_d^2 - \br_{\bx d}\Th\hat\bw)+\tnr{Const.}
\end{align}

Using the eigenvalue decomposition,
$\matR_{\bx} = \matQ\tnr{diag}(\blambda)\matQ\Th$, 
where $\tnr{diag}(\blambda)$ is a diagonal matrix with diagonal elements taken from the vector $\blambda=[\lambda_1,\ld,\lambda_L]\T$, $\lambda_l$ being the eigenvalues of $\matR_{\bx}$, and the columns of $\matQ\in\Real^{L\times L}$ are the corresponding eigenvectors, 
we obtain
\begin{align}
\hat\bw(\alpha) &= \matQ\tnr{diag}^{-1}(\blambda + \alpha)\bz_{\bx d},\\
\label{bz.bx.d}
\bz_{\bx d} & = \matQ\Th\br_{\bx d} = [z_{\bx d,1},\ld, z_{\bx d, M}]\Th,
\end{align}
so \eqref{L(alpha)} may be written as 
\begin{align}
\label{L(alpha).eigenvalues}
    L(\alpha)
    &=
    \sum_{m=1}^M\log\frac{\alpha+\lambda_m}{\alpha} + N\log\left(\tilde\sigma_d^2 - \sum_{m=1}^M\frac{z_{\bx d,m}^2}{\alpha+\lambda_m}\right) +\tnr{Const} ,
\end{align}
and, now, we easily find its derivative:
\begin{align}
\label{L.prime.f.g.gamma}
L'(\alpha)
&=N \frac{f(\alpha)}{g(\alpha)}
-\frac{\gamma(\alpha)}{\alpha},
\end{align}
where 
\begin{align}
\label{f(alpha)}
f(\alpha)& = 
\sum_{m=1}^M \frac{z_{\bx d, m}^2}{(\lambda_m+\alpha)^2}
= \gamma(\alpha) \tilde\sigma_w^2(\alpha),\\
\label{g(alpha)}
g(\alpha) 
&=\tilde\sigma_d^2-\br_{\bx d}\Th\hat{\bw}(\alpha) 
= \tilde\sigma_e^2(\alpha) + \alpha \gamma(\alpha) \tilde\sigma_w^2(\alpha),
\end{align}
in which $\tilde\sigma_e^2(\alpha)$ and $\tilde\sigma_w^2(\alpha)$ are given in \eqref{hat.sigma.e.HKL} and \eqref{hat.sigma.w.HKL}, respectively, and the latter uses 
\begin{align}
\label{gamma(alpha)}
\gamma \equiv \gamma(\alpha) &= \sum_{m=1}^M\frac{\lambda_m}{\lambda_m+\alpha},
\end{align}
also known as the effective number of parameters \cite[Sec.~7.6]{Hastie_book}. Note that $\gamma(\alpha) \in [0, M]$ and, for $\alpha=0$, if no eigenvalues are zero, we can use $\gamma=\gamma(\alpha)=M$, as we did in \eqref{hat.sigma.w.HKL}.


As already noted in \cite{Selen08}, solving $L'(\alpha)=0$ amounts to finding the real roots of the polynomial of degree not larger than $2M-1$, whose properties are described in the following:

\begin{proposition}[Roots of $L'(\alpha)$]
\label{Proposition:existence.finite.regularization}
~

\begin{enumerate}
\item 
    $\lim_{\alpha\rightarrow\infty}L(\alpha)=0$, \ie $\alpha=\infty$ is a root of $L'(\alpha)$.
\item 
    The odd-numbered roots (the first, the third, etc.) of $L'(\alpha)$ are minima of $L(\alpha)$.
\item 
    $L'(\alpha)$ has an even number of roots if and only if 
\begin{align}
\label{Condition.for.finite.alpha}
N\|\br_{\bx d}\|^2>\tilde\sigma_d^2\tnr{Tr}(\matR_{\bx}).
\end{align}
\end{enumerate}

\textbf{Proof:} \ref{App:Proof.proposition.existence} 
\end{proposition}

Some comments are in order.
\begin{itemize}
\item We should appreciate the possibility of absence of finite roots of $L'(\alpha)$. Note that, if the only root\footnote{Of course, we talk about the real roots which are meaningful solutions.} is $\alpha=\infty$, then it is also the first root, which means that $L(\alpha)$ is minimized for $\alpha=\infty$, in which case $\hat\bw(\alpha)=\bzero$. The fact that such a solution may be optimal is not at all obvious when formulating the filtering problem. As we will see empirically, it is indeed the case in some scenarios.

\item Since $\tilde\sigma_d$, $\br_{\bx d}$, and $\matR_{\bx}$ are empirical means, which, for large $N$, tend to its corresponding expected values, \eqref{Condition.for.finite.alpha} is likely to be satisfied for sufficiently large $N$, where the latter dominates the \gls{lhs} of \eqref{Condition.for.finite.alpha}. In other words, by increasing $N$, we will have an even number of roots and then $\alpha=\infty$ is a local maximum of $L(\alpha)$ and thus $\alpha_{\tnr{ML}}$ is finite.

\end{itemize}

Finding the roots may be done exploiting the polynomial structure of $L'(\alpha)$ but, in practice, this is feasible only for moderate $M$, \eg in \gls{mvdr} receivers applied in arrays composed of dozens of antennas. For large $M$, \eg $M>100$, typical in system identification and/or equalization, the roots may be found, \eg via grid search \cite{Selen08}. However, not all of these methods are very practical, which may explain why they did not receive much attention in the literature -- in fact, they were not reused as a go-to-solution by the authors of \cite{Selen08}, \eg in \cite{Du10}.

Our goal is thus to propose a simple approach to solve $L'(\alpha)=0$, which, after reorganizing \eqref{L.prime.f.g.gamma}, is equivalent to solving 
\begin{align}
\label{alpha=gamma.g.f}
\alpha &=\gamma(\alpha)\frac{g(\alpha)}{Nf(\alpha)}, 
\end{align}
which we do via a fixed-point iteration:
\begin{align}
\label{alpha.i+1.i}
\alpha^{(i+1)} &= \gamma\big(\alpha^{(i)}\big)\frac{g\big(\alpha^{(i)}\big)}{Nf\big(\alpha^{(i)}\big)},\\
\label{alpha.i+1.i.new}
&=\frac{\tilde\sigma_e^2\big(\alpha^{(i)}\big)}{N \tilde\sigma_w^2\big(\alpha^{(i)}\big)} + \frac{\alpha^{(i)}}{N}\gamma\big(\alpha^{(i)}\big) ,\quad i=1,\ld, I,\\
\alpha_{\tnr{ML}} & = \alpha^{(I)},
\end{align}
where $I$ is a predefined number of iterations, and initialization $\alpha^{(0)}>0$ must be defined.

Note that:
\begin{itemize}

\item The convergence of the fixed-point iteration \eqref{alpha.i+1.i} is not proven, but, in numerical examples, it always converged to a
minima of $L'(\alpha)$ when \eqref{Condition.for.finite.alpha} was satisfied (\ie when there are finite minima of $L(\alpha)$).

\item With the initialization $\alpha^{(0)}=0$, the first iteration of \eqref{alpha.i+1.i.new} yields
\begin{align}
    \alpha^{(1)} & = \frac{\tilde{\sigma}_e^2(0)}{N\tilde{\sigma}_w^2(0)},
\end{align}
which is exactly the HKB method shown in \eqref{alpha.HKB}. We can thus say that our solution generalizes the HKB method, enhancing it with an iterative refinement, and removing the initialization with a non-regularized solution, \ie $\alpha^{(0)}=0$, which may be problematic in general, since it cannot be solved meaningfully for $N<M$. 

\item The fixed-point iteration \eqref{alpha.i+1.i.new} is not a unique way to solve the problem iteratively. For example, using \eqref{g(alpha)} in \eqref{alpha=gamma.g.f}, and isolating $\alpha$, we obtain a new fixed-point equation:
\begin{align}
\label{alpha.Gull.McKay}
\alpha^{(i+1)}
&= \frac{\tilde\sigma_e^2\big(\alpha^{(i)}\big)}{[N- \gamma\big(\alpha^{(i)}\big)] \tilde\sigma_w^2\big(\alpha^{(i)}\big)},
\end{align}
which is known as Gull-MacKay iteration \cite[Ch.~18.1.4]{Barber12_Book}; see \cite[v1~App.~A]{Zanco23b} for an alternative derivation. 

Our experience shows that Gull-MacKay converges faster than \eqref{alpha.i+1.i}. However, it should be applied with care for $N<M$, because in this case, we do not have a guarantee that $N-\gamma(\alpha)$ is positive [as seen in \eqref{gamma(alpha)}].

\item The iterative solutions \eqref{alpha.i+1.i.new} and \eqref{alpha.Gull.McKay} use $\tilde\sigma_w^2(\alpha)$ and $\tilde\sigma_e^2(\alpha)$ which may be calculated from the eigenvalue decomposition shown in \eqref{L(alpha).eigenvalues}-\eqref{gamma(alpha)}; this reduces the complexity significantly.
\end{itemize}

\section{Automatic regularization in the interference-suppression problem}\label{Sec:Regularization.for.MVDR}

Having solved the problem of automatic regularization of the Wiener equations \eqref{hat.h} in the error-minimization problem, we turn our attention to the interference-suppression problem \eqref{MVDR.problem.definition} and we reformulate it to take advantage of the development we already made in Sec.~\ref{Sec:find.alpha}. To this end, we need to remove the constraint in \eqref{MVDR.problem.definition}, which is done by expressing $\bw$ as
\begin{align}\label{bw.M.A.u}
\bw = \frac{1}{M} \ba - \matA\bu,
\end{align}
where
\begin{align}
\matA &= \matI - \frac{1}{M} \ba\ba\Th = \matA\Th
\end{align}
is the projection matrix; indeed, it is easy to see that $\ba\Th\matA=\bzero$, and thus, for any $\bu\in\Complex^M$, $\ba\Th\bw=1$.

Note that this approach, with a slightly different definition of \eqref{bw.M.A.u}, was also used in \cite{Selen08,Du10}.

We may thus reformulate \eqref{MVDR.problem.definition} as
\begin{align}
\label{from.h.to.w}
\hat\bw &= \frac{1}{M}\ba-\matA\hat\bu,\\
\label{hat.bh.MVDR}
\hat\bu &=\argmin_{\bu} \left\{ \Ex\Big[|(\ba/M-\matA\bu)\Th\bx(t)|^2\Big] +\alpha \|\ba/M-\matA\bu\|^2 \right\},\\
\label{argmin.h}
&=\argmin_{\bu} \left\{ \Ex\Big[|\tilde{d}(t) - \bu\Th\tilde\bx(t)|^2\Big] + \alpha \bu\Th\matA\bu \right\},
\end{align}
where we removed the constant terms from \eqref{argmin.h}, and we used
\begin{align}
\tilde\bx(t) &= \matA\bx(t),\\
\tilde{d}(t) &= \frac{1}{M} \ba\Th\bx(t).
\end{align}

\begin{proposition}\label{Proposition.h.opt.equivalent}
The optimization in \eqref{argmin.h} is equivalent to 
\begin{align}
\label{argmin.h.reformulated}
\hat\bu
&=\argmin_{\bu} \left\{ \Ex\Big[|\tilde{d}(t) - \bu\Th\tilde\bx(t)|^2\Big] + \alpha \|\bu\|^2 \right\} .
\end{align}

\textbf{Proof}:
We can always write $\bu=\bu_{\parallel}  + \bu_{\perp}$, where $\bu_{\parallel}=\beta\ba$ is the term collinear with $\ba$ and $\bu_{\perp}$ is the term orthogonal to $\ba$, \ie $\ba\Th\bu_{\perp}=0$. Then,  from $\bu_{\parallel}\Th\matA=0$, we see that $\bu\Th\tilde\bx(t)=\bu_{\perp}\Th\tilde\bx(t)$ and $\bu\Th\matA\bu=\|\bu_{\perp}\Th\|^2$ , which means that the cost function under minimization in \eqref{argmin.h} is insensitive to adding a term collinear with $\ba$ to any $\bu$, \ie $\bu+\beta\ba$. In particular, we may remove the term collinear with $\ba$ from $\hat\bu$ by adding a penalty term $\|\bu_{\parallel}\|^2$, \ie
\begin{align}
\label{hat.u.new}
\hat\bu
&=\argmin_{\bu} \left\{ \Ex\Big[|\tilde{d}(t) - \bu\Th\tilde\bx(t)|^2\Big] + \alpha ( \bu\Th\matA\bu + \|\bu_{\parallel}\|^2) \right\},
\end{align}
and, because $\bu\Th\matA\bu + \|\bu_{\parallel}\|^2=\|\bu\|^2$, \eqref{hat.u.new} is the same as \eqref{argmin.h.reformulated}.
\end{proposition}

The goal of Proposition~\ref{Proposition.h.opt.equivalent} was to obtain \eqref{argmin.h.reformulated} which has the same form as error-minimization problem \eqref{hat.h}. Thus, we can reuse the equations of the latter, \ie 
\begin{align}
\hat\bu & = (\matR_{\tilde\bx}+\alpha\matI)^{-1}\br_{\tilde\bx \tilde{d}},\\
\matR_{\tilde\bx} &= \frac{1}{N}\sum_{t=0}^{N-1}\tilde\bx(t)\tilde\bx\Th(t)=\matA\matR_{\bx}\matA,\\
\br_{\tilde\bx \tilde{d}} &= \frac{1}{N}\sum_{t=0}^{N-1}\tilde\bx(t)\tilde{d}^*(t)= \frac{1}{M}\matA\matR_{\bx}\ba,
\end{align}
as well as we can apply the iterative solution \eqref{alpha.i+1.i.new} to find the regularization factor, that is 
\begin{align}
\label{alpha.i+1.i.MVDR}
\alpha^{(i+1)} &= \gamma\big(\alpha^{(i)}\big) \frac{\frac{1}{N}\| \tilde\bd - \tilde\matX\Th\hat\bu\big(\alpha^{(i)}\big)\|^2}{N \|\hat\bu\big(\alpha^{(i)}\big)\|^2} + \frac{\gamma\big(\alpha^{(i)}\big)}{N} \alpha^{(i)}
\end{align}


Since we removed the terms collinear with $\ba$, we have $\hat\bu\Th\ba=0$, and
\begin{align}
\|\hat\bu^{(i)}\|^2 = \|\hat\bw^{(i)}\|^2-\frac{1}{M},
\end{align}
and, from \ref{sec:gamma}, we have
\begin{align}
\label{gamma.i.from.appendix}
\gamma^{(i)} = M - \alpha^{(i)} \tnr{Tr}[(\matR_{\bx} + \alpha^{(i)}\matI)^{-1}] -  \ba\Th \matR_{\bx} \Th(\matR_{\bx} + \alpha^{(i)}\matI)^{-1}\hat\bw^{(i)},
\end{align}
which may be integrated in the fixed point iteration.

For example, the Gull-MacKay iteration \eqref{alpha.Gull.McKay} becomes 
\begin{align}
\label{hat.w.i}
\hat\bw^{(i)} &= \frac{(\matR_{\bx} + \alpha^{(i)}\matI)^{-1}\ba}{\ba\Th(\matR_{\bx} + \alpha^{(i)}\matI)^{-1}\ba},\\
\label{alpha.i+1.Gull-MacKay.MVDR}
\alpha^{(i+1)} &= \frac{(\hat\bw^{(i)})\Th \matR_{\bx} \hat\bw^{(i)}}{\left(\|\hat\bw^{(i)}\|^2 - \displaystyle\frac{1}{M} \right) \left( \displaystyle\frac{N}{\gamma^{(i)}} - 1 \right)}.
\end{align}

\section{Numerical examples}\label{Sec:Numerical.examples}

\subsection{Error-minimization problem: system identification}\label{Sec:Numerical.identification}

\begin{figure}[h!]
\centering
\subfloat[]{\includegraphics[width=0.6\textwidth]{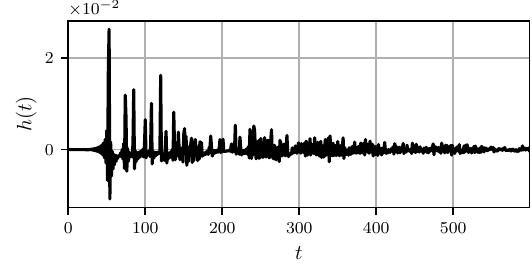}}\\
\caption{Impulse response $\bh$ 
generated using \cite{Werner2023}.}
\label{fig:impulses}
\end{figure}

We consider the problem of identification of an acoustic impulse response, where $x(t)$ is an AR(1) process, \ie $x(t)=ax(t-1)+u(t)$ and $u(t)$ is generated from a zero-mean unit-variance white Gaussian noise; we use $a=0.9$. The impulse response $\bh=[h(0),\ld, h(M-1)]\T$ with length $M=600$, shown in \figref{fig:impulses}, is calculated using software \cite{Werner2023} for a room of dimensions $(5, 4, 6)$~m, the source in position $(2, 3.5, 2)$~m, the receiver in position $(2, 1.5, 1)$~m, a sampling rate of $8$~kHz, and a reverberation time of $225$~ms. The desired output is obtained as $d(t)=\bh\T\bx(t)+e(t)$, with $e(t)$ being a zero-mean Gaussian noise with variance $v^*_e$, and
\begin{align}
    \bx(t) = [x(t), x(t-1),\dots, x(t-M+1)]\T.
\end{align}

We define the \gls{snr} as
\begin{align}
\tnr{SNR} & = 10\log_{10}\left(\frac{\Ex[|\bh\T\bx(t)|^2]}{v^*_e}\right)~[\tnr{dB}].
\end{align}

Although we use real variables, it is easy to see that the formulas to find $\alpha$, derived in Sec.~\ref{Sec:find.alpha}, are the same.

The quality of the estimate $\hat\bw\equiv \hat\bw(\alpha)$ will be assessed through the misalignment (a relative estimation error) of the impulse response:
\begin{align}    
\label{Misalignement.def}
\mfm(\alpha) &= 20\log_{10}\left( \frac{\|\hat\bw - \bh\|_2}{\|\bh\|_2} \right) [\tnr{dB}].
\end{align}

A simple, worst-case metric, is obtained by setting $\alpha=\infty$, for which $\hat\bw=\bzero$, and thus we have $\mfm(\infty)=0~\tnr{dB}$. The best-case reference is obtained with ``oracle"-given regularization parameter and its corresponding misalignment:
\begin{align} \label{hat.alpha}
\hat\alpha &= \argmin_{\alpha}\mfm(\alpha),\\
\label{hat.M}
\hat{\mfm} & = \mfm(\hat\alpha).
\end{align}

\begin{figure}
\centering
\includegraphics{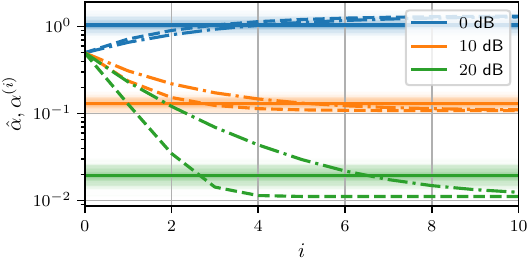}
\caption{Values $\alpha^{(i)}$ obtained via fixed-point iteration \eqref{alpha.i+1.i.new} (dashed-dotted lines) and via Gull-MacKay iteration \eqref{alpha.Gull.McKay} (dashed lines) in different realizations of the data using $\alpha^{(0)}=0.5$; $N = 1000$, $M = 600$. Solid lines are constant, as they indicate an oracle-given $\hat\alpha$. Thick lines indicate averages over realizations shown with thin lines.}
\label{fig:fixed-point.iter}
\end{figure}

Fig.~\ref{fig:fixed-point.iter} illustrates the convergence of fixed point iterations \eqref{alpha.i+1.i.new} and \eqref{alpha.Gull.McKay}: it shows the evolution of $\alpha^{(i)}$ with the starting point $\alpha^{(0)}=0.5$, chosen to be far from the oracle-given $\hat\alpha$. We evaluate various realizations of the data with $N=1000$ and $M=600$, and note that, beyond $I=5$, for practical purposes, convergence may be declared for Gull-MacKay, while the fixed-point iteration \eqref{alpha.i+1.i.new} is slower, requiring approximately twice as many iterations.

All the results we show in the following are thus based on the Gull-MacKay iteration, with $I=5$ and $\alpha^{(0)} = 0.5$. We verified that, in all displayed cases, the condition \eqref{Condition.for.finite.alpha} was not violated.\footnote{This is because we decided to use $N>M$ which is a practical approach to the system identification. However, for smaller $N$, the condition \eqref{Condition.for.finite.alpha} may be violated.}

\begin{figure}[ht!]
\centering
\subfloat[]{\includegraphics[width=.49\textwidth]{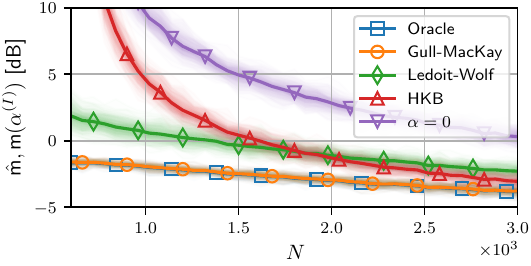}} \, \subfloat[]{\includegraphics[width=.49\textwidth]{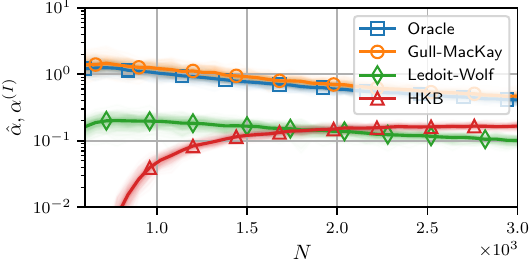}}\\
\subfloat[]{\includegraphics[width=.49\textwidth]{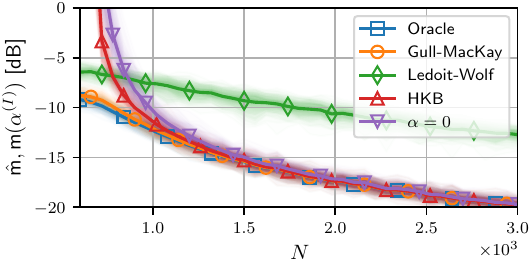}} \, \subfloat[]{\includegraphics[width=.49\textwidth]{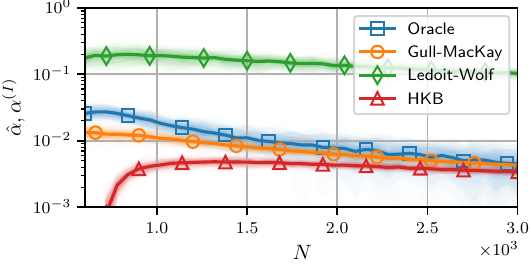}}\\
\caption{Results obtained for different regularization methods, and SNR equal to (a,b) $0$dB and (c,d) $20$dB. In (a,c) we show the misalignment $\mfm(\alpha^{(I)})$ \eqref{Misalignement.def}, and $\hat{\mfm}$ given by \eqref{hat.M}, while the corresponding values of $\alpha^{(I)}$ and $\hat\alpha$ are shown in (b,d). In (b) and (d), thick lines are averages of realizations shown with thin lines. 
}
\label{fig:best_alpha}
\end{figure}

The results, shown in Fig.~\ref{fig:best_alpha}(a)(c), 
are consistent with intuition: by increasing $N$ and $\tnr{SNR}$, we decrease the estimation error when the oracle and the fixed-point (Gull-MacKay) iteration regularization is used. In fact, the difference between the regularization parameter $\alpha^{(I)}$ and the oracle-given value $\hat\alpha$ is rather small, making the iterative estimation \eqref{alpha.Gull.McKay} an attractive tool for the choice of $\alpha$.

Moreover, we observe that 
(i) the HKB and the Ledoit-Wolf regularization methods may yield worse performance than $\mfm(\infty)=0$~dB, which is the trivial performance limit. This is well understood for $N<M$, because then $\alpha_{\tnr{HKB}}=0$, \ie the solution is not regularized; see our comments at the end of Sec.~\ref{Sec:HKB}. Moreover, for low \gls{snr}, the HKB regularization requires a substantial number of samples (approx. $N>1600$) to merely attain $\mfm(\alpha) = 0~\tnr{dB}$,
(ii) the Ledoit-Wolf regularization does not adapt to the data, \eg for large \gls{snr} it fails to outperform the non-regularized ($\alpha=0$) solution. This is not entirely surprising because the Ledoit-Wolf method does not take into account the cross-correlation $\br_{\bx d}$.

\subsection{Interference suppression problem: beamforming}\label{Sec:Numerical.MVDR}

We consider the antenna-processing scenario, in which the signal $\bx(t)$ \eqref{x.t.MVDR} is defined as
\begin{align}
\bx(t) = \sum_{k=1}^K d_k(t) \ba(\phi_k) + \be(t),
\end{align}
where $\be(t)$ is a zero-mean, circular complex Gaussian noise with covariance matrix $\Ex[\be(t)\be\Th(t)]=\matI$, and $d_k(t)$ are zero-mean, unit-variance, \gls{iid} Gaussian variables modeling signals, each with power $\sigma_k^2=\Ex[|d_k(t)|^2]$, and the steering vector for the angle $\phi$ is defined as 
\begin{align}
\ba(\phi) &= [1, \e^{-j \pi \cos(\phi)}, \e^{-j 2\pi \cos(\phi)},\dots, \e^{-j(M-1) \pi \cos(\phi)}]\T,
\end{align}
that is, we assume that $\bx(t)$ is acquired at a linear antenna array with $M$ elements spaced at half-wavelength \cite[Ch.~6.5]{Sayed08_Book}.

The true covariance matrix is thus calculated as
\begin{align}
\ov\matR_{\bx} &= \sum_{k=1}^K \sigma_k^2 \ba(\phi_k)\ba(\phi_k)\Th + \matI.
\end{align}

In the beamforming problem, our goal is to suppress the interference signals $d_l(t), l\neq k$ using the filter $\hat\bw_k$ found through \eqref{hat.w.MVDR}, where we know the steering vector of the signal of interest $\ba=\ba(\phi_k), k \in \set{1,\dots, K}$. The quality of interference suppression is measured by the \gls{sinr} at the output of the filter, calculated as
\begin{align}
\label{SINR.opt}
\mf{SINR}_k &= 
\frac{\Ex[|d_k(t)|^2]}
{\Ex[|\hat\bw_k\Th\bx(t)|^2] - \Ex[|d_k(t)|^2]} = 
\frac{\sigma^2_k}
{\hat\bw_k\Th\ov\matR_{\bx}\hat\bw_k - \sigma^2_k}.
\end{align}



In this example, we use $K=3$, $[\sigma^2_1, \sigma^2_2, \sigma^2_3]=[20, 10, 5] \tnr{dB}$, and $[\phi_1,\phi_2, \phi_3]= [0.2\pi, 0.3\pi, 0.6\pi]$. 

\begin{figure}
\centering
\includegraphics[]{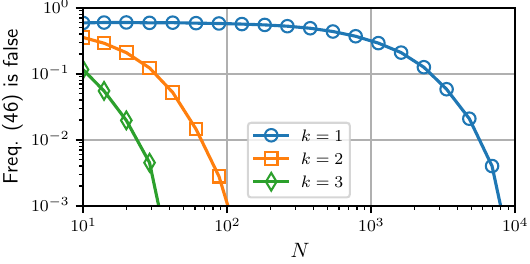}
\caption{Empirical frequency of violating condition \eqref{Condition.for.finite.alpha} in the interference suppression example with 10000 independent realizations.}
\label{fig:capon-condition}
\end{figure}

We show in Fig.~\ref{fig:capon-condition} the empirical frequency of violating condition \eqref{Condition.for.finite.alpha} obtained from 10000 data realizations. In these cases, $L'(\alpha)$ often has no finite roots, \ie $\alpha_{\tnr{ML}}=\infty$, $\hat\bu=\bzero$ and $\hat\bw = \ba/M$. In other words, there are cases where the optimal solution $\hat\bw$ is a matched filter.

To understand why this may happen, we recall that the matched filter is optimal in the presence of white Gaussian noise. This clarifies why the probability of obtaining such a solution is larger for high-energy target signal (\eg $k=1$): this is when the interference is weak and may, indeed, ``appear like'' white noise, especially for small $N$. On the other hand, for weak signals (\eg $k=3$), the interference (\eg from the signal $k=1$) is strong and will emerge from the empirical covariance matrix $\matR_{\bx}$, even for relatively small $N$.


\begin{figure}[ht!]
\centering
\subfloat[]{\includegraphics[width=0.48\textwidth]{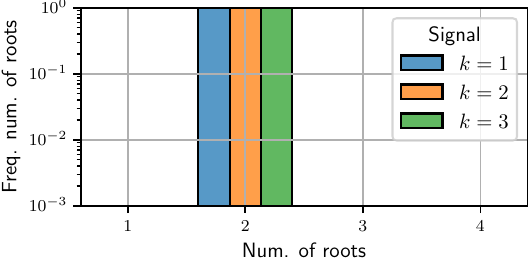}}\,
\subfloat[]{\includegraphics[width=0.48\textwidth]{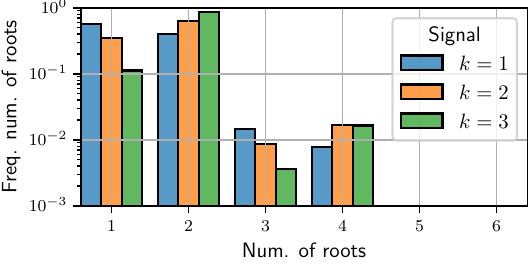}}
\caption{Empirical frequency of the number of roots in the interference suppression example across different signals of interest for 10000 independent realizations of data, $M=10$ and (a) $N = 10000$, (b) $N = 10$.}
\label{fig:capon-num_roots}
\end{figure}

The empirical evaluation of the number of roots of $L'(\alpha)$, is shown in Fig.~\ref{fig:capon-num_roots} for large and small number of samples $N$, leads to the following observations: (i) for large $N$, the vast majority of cases produced a unique and finite root $\alpha_{\tnr{ML}}$, which was obtained here through the Gull-MacKay iteration \eqref{alpha.i+1.Gull-MacKay.MVDR} (since there are two roots, the first one is the minimum, see Proposition~\ref{Proposition:existence.finite.regularization}b); (ii) for small $N$, frequent cases are when $\alpha_{\tnr{ML}}=\infty$ (when there is one root) or when there are multiple finite roots; it occurs relatively frequently, especially for strong target signals $k\in\set{1,2}$; (iii) in the presence of multiple minima, the matched filter solution $\alpha=\infty$ can be competitive with $\alpha_{\tnr{ML}}$, \ie $L(\alpha_{\tnr{ML}}) \approx L(\infty)$.

To handle the multiple-roots situation, without explicitly identifying them all (which may be numerically tedious), we propose a two-step approach: First, we find the root $\alpha_{\tnr{ML}}$ using the Gull-MacKay iteration \eqref{alpha.i+1.Gull-MacKay.MVDR}. Next, we verify if $L(\alpha_{\tnr{ML}}) > L(\infty)$, in which case we make a replacement $\alpha_{\tnr{ML}}\leftarrow\infty$, otherwise we keep $\alpha_{\tnr{ML}}$ unchanged. In fact, this heuristic is easy to implement because, from \eqref{L(alpha).eigenvalues}, we have $L(\infty) = N \log \tilde\sigma_d^2$.

\begin{figure}[ht!]
\centering
\subfloat[]{\includegraphics{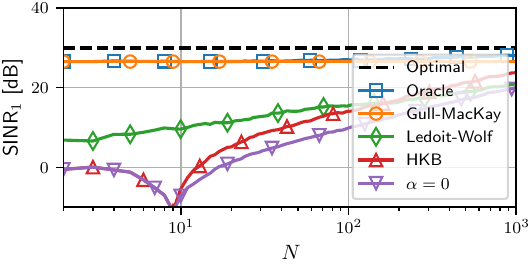}}\\
\subfloat[]{\includegraphics{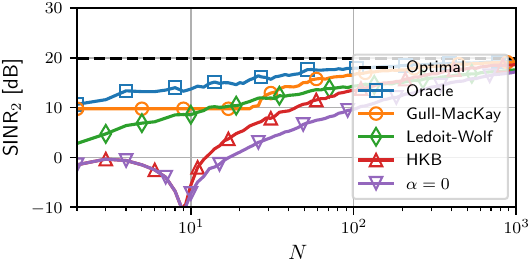}}\\
\subfloat[]{\includegraphics{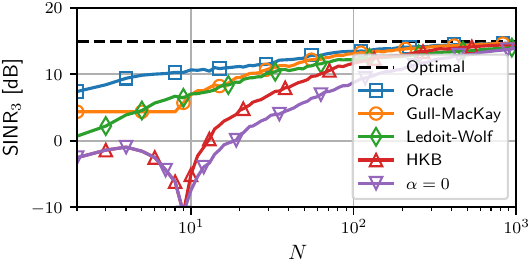}}\\
\caption{\gls{sinr} \eqref{SINR.opt} obtained for a) $k=1$, b) $k=2$, and c) $k=3$. }
\label{fig:SINR.vs.N}
\end{figure}

In Fig.~\ref{fig:SINR.vs.N}, we show $\mf{SINR}_k, k=1,2,3$ as a function of $N$, for different regularization methods. 

\begin{figure}[ht!]
\centering
\subfloat[]{\includegraphics{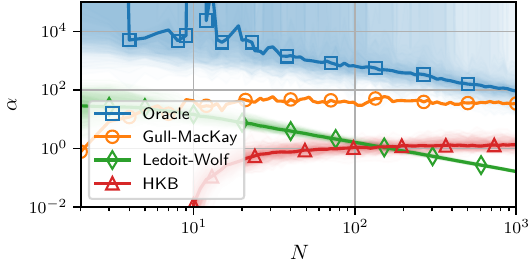}}\\
\subfloat[]{\includegraphics{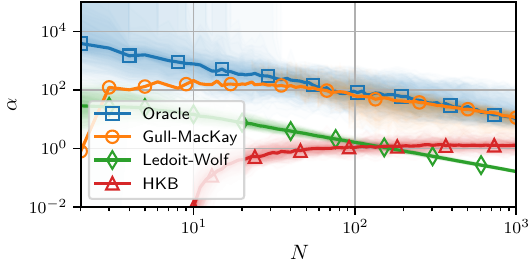}}\\
\subfloat[]{\includegraphics{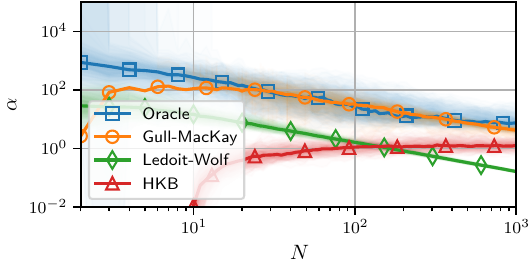}}\\
\caption{The regularization parameter obtained for a) $k=1$, b) $k=2$, and c) $k=3$.}
\label{fig:alpha.vs.N}
\end{figure}

Similarly, the values of the regularization parameter are shown in Fig.~\ref{fig:alpha.vs.N}. In this case, the thick line corresponds to the median of the regularization parameter, as it gracefully deals with the cases when $\alpha_{\tnr{ML}}=\infty$.

We observe that (i) the proposed estimation method is very close to the oracle solutions, and clearly outperforms other methods, especially for $N>M$ and for strong target signal $k=1$, (ii) In many cases, for relatively small $N$ and high target signal power ($k=1$), the optimal regularization is $\alpha_{\tnr{ML}}=\infty$, which means that the optimal solution is a matched filter, see \eqref{hat.w.i}, (iii) as in Sec.~\ref{Sec:Numerical.identification}, the HKB regularization approaches the optimal solution only for sufficiently large $N$, and (iv) the Ledoit-Wolf regularization parameters is independent of the steering vector $\ba$ (see Fig.~\ref{fig:alpha.vs.N}) which affects its performance; this illustrates well the idea that, in the \gls{mvdr} problem, the regularization should take into account the steering vector and not only the covariance matrix $\matR_{\bx}$.

\section{Conclusions}
In this work, we presented a method, adopted from the area of statistical machine learning, to find the regularization parameter in two main classes of linear \gls{mmse} filters applied in  (i) the error-minimization and (ii) the interference suppression problems. Using a probabilistic formulation, we estimate the parameters of the model from the \gls{ml} principle, where the regularization parameter is found using a few steps of the fixed-point iteration. We also provide data-dependent conditions for the existence of the finite \gls{ml} solution and show heuristics which deal well with multiple \gls{ml} solutions.

Numerical examples indicate that the simple iterative solution we show is remarkably close to the optimal regularization parameter.

We compare the proposed solution with other methods known in the literature. We show that the HKB method \cite{ArthurE.Hoerl1975} may be seen as a simplified version of our approach and that the Ledoit-Wolf shrinkage \cite{Ledoit04} fails to appropriately choose the regularization, which is due to its explicit independence from the desired signal.

\section*{Acknowledgments}

This work was supported in part by the \textit{Fonds de recherche du Québec} (FRQ) - \textit{Nature et technologies} under the \textit{Doctoral research scholaships B2X 2024-2025} program, file number 342496, recipient Daniel~Gomes~de~Pinho~Zanco.

\appendix
\section{Proof of Proposition~\ref{Proposition:existence.finite.regularization}}\label{App:Proof.proposition.existence}

Considering \eqref{L.prime.f.g.gamma}, we note that $f(\alpha)$ and $g(\alpha)$ shown in \eqref{f(alpha)} and \eqref{g(alpha)} are bounded and positive, therefore, their ratio is also bounded and positive. 

Since $\lim_{\alpha\rightarrow 0 }\gamma(\alpha)/\alpha = \infty$, for a sufficiently small $\alpha$, we have $L'(\alpha)<0$ (\ie $\exists\alpha^*,\forall\alpha<\alpha^*,  L'(\alpha)<0$). We also have that
\begin{equation}
\label{eq:infinity.root}
\begin{aligned}
\lim_{\alpha\rightarrow\infty} L'(\alpha) &= \lim_{\alpha\rightarrow\infty} \frac{N f(\alpha) \alpha - g(\alpha) \gamma(\alpha)}{\alpha g(\alpha)}\\
&= \lim_{\alpha\rightarrow\infty} \frac{N \alpha \displaystyle\sum_{m=1}^M \frac{z_{\bx d, m}^2}{(\lambda_m+\alpha)^2} - \left[\tilde\sigma_d^2- \displaystyle\sum_{m=1}^M \frac{z_{\bx d, m}^2}{\lambda_m+\alpha} \right] \left[\displaystyle\sum_{m=1}^M \frac{\lambda_m}{\lambda_m+\alpha} \right]}{\alpha \left[ \tilde\sigma_d^2 -\displaystyle\sum_{m=1}^M \frac{z_{\bx d, m}^2}{\lambda_m+\alpha} \right]},\\
&= \frac{\displaystyle\lim_{\alpha\rightarrow\infty} N\displaystyle\sum_{m=1}^M \frac{z_{\bx d, m}^2}{(\lambda_m+\alpha)^2} - \displaystyle\frac{1}{\alpha}\left[\tilde\sigma_d^2- \displaystyle\sum_{m=1}^M \frac{z_{\bx d, m}^2}{\lambda_m+\alpha} \right] \left[\displaystyle\sum_{m=1}^M \frac{\lambda_m}{\lambda_m+\alpha} \right]}{\tilde\sigma_d^2 - \displaystyle\lim_{\alpha\rightarrow\infty} \displaystyle\sum_{m=1}^M \frac{z_{\bx d, m}^2}{\lambda_m+\alpha}},\\
&= 0,
\end{aligned}
\end{equation}
thus $\infty$ is a root of $L'(\alpha)$.


Then, from intermediate value theorem, $L'(\alpha)$ has at least one finite root if $L'(\alpha) > 0$ for a sufficiently large $\alpha$, \ie when 
\begin{align}
N \frac{\alpha f(\alpha)}{\gamma(\alpha) g(\alpha)} > 1.
\end{align}
By taking the limit as $\alpha$ tends to $\infty$ on both sides, we can evaluate if $L'(\alpha)$ is decreasing, such that if
\begin{align}
N \lim_{\alpha\rightarrow \infty} \frac{\alpha f(\alpha)}{g(\alpha)\gamma(\alpha)} &> 1 \\
N \lim_{\alpha\rightarrow \infty} \frac{\alpha f(\alpha)}{\gamma(\alpha)} &> \lim_{\alpha\rightarrow \infty} g(\alpha)\\
N \lim_{\alpha\rightarrow\infty} \frac{\alpha \displaystyle\sum_{m=1}^M \frac{z_{\bx d, m}^2}{(\lambda_m+\alpha)^2}}{\displaystyle\sum_{m=1}^M \frac{\lambda_m}{\lambda_m+\alpha}} &> \tilde\sigma_d^2\\
N \lim_{\alpha\rightarrow\infty} \frac{\displaystyle\sum_{m=1}^M \frac{z_{\bx d, m}^2}{(\frac{\lambda_m}{\alpha}+1)^2}}{\displaystyle\sum_{m=1}^M \frac{\lambda_m}{\frac{\lambda_m}{\alpha}+1}} &> \tilde\sigma_d^2\\
N \frac{\displaystyle\sum_{m=1}^M z_{\bx d, m}^2}{\displaystyle\sum_{m=1}^M \lambda_m} &> \tilde\sigma_d^2\\
\label{eq:fp.condition}
N\|\br_{\bx d}\|^2>\tilde\sigma_d^2\sum_{m=1}^M\lambda_m,
\end{align}
where \eqref{eq:fp.condition} is the same as \eqref{Condition.for.finite.alpha}. 


When \eqref{eq:fp.condition} is true, $L'(\alpha)$ changes sign at least once, and thus $L'(\alpha)$ has at least 2 roots (one at $\infty$ and the other at the sign change). If there are 3 roots, then \eqref{eq:fp.condition} cannot be true, since $L'(0) < 0$ and $L'(\infty) = 0$, and three roots would require two sign changes. These observations can be extended to an arbitrary number of roots. In fact, \eqref{eq:fp.condition} can only be true if the number of roots is even, and, since $\infty$ is always a root, the condition also tells us if there is at least one finite root.

This finishes the proof.



\section{Derivation of $\gamma$ in MVDR filter}\label{sec:gamma}

To calculate $\gamma=\tnr{Tr}[\matI-\alpha(\matA\matR_{\bx}\matA + \alpha\matI)^{-1}]$, we find
\begin{align}
\matI - \alpha(\matA\matR_{\bx}\matA + \alpha\matI)^{-1}
&= \matA(\alpha\matR_{\bx}^{-1} + \matI -\ba\ba\Th/M )^{-1}\matA,\\ 
&=
\matA[(\matR_{\bx} + \alpha\matI)\matR_{\bx}^{-1} - \ba\ba\Th/M ]^{-1}\matA,\\
&=
\matA\left[\matR_{\bx}(\matR_{\bx} + \alpha\matI)^{-1} + \frac{\matR_{\bx}(\matR_{\bx} + \alpha\matI)^{-1} \ba\ba\Th \matR_{\bx}(\matR_{\bx} + \alpha\matI)^{-1}}{M - \ba\Th \matR_{\bx}(\matR_{\bx} + \alpha\matI)^{-1}\ba} \right]\matA,\\
&=
\matA\left[\matS + \frac{\matS \ba\ba\Th \matS}{M - \ba\Th \matS\ba} \right]\matA,
\end{align}
where $\matS = \matR_{\bx}(\matR_{\bx} + \alpha\matI)^{-1} = \matI - \alpha(\matR_{\bx} + \alpha\matI)^{-1}$. 

Next, using the fact that $\matA \matA = \matA$,
\begin{align}
\gamma &= \tnr{Tr}(\matS\matA) + \frac{\tnr{Tr}(\matS \ba\ba\Th \matS\matA)}{M - \ba\Th \matS\ba},\\
&= \tnr{Tr}(\matS) - \frac{1}{M} \ba\Th\matS\ba + \frac{\ba\Th\matS\matS\ba - \frac{1}{M} |\ba\Th\matS\ba|^2}{M - \ba\Th \matS\ba},\\
&= \tnr{Tr}(\matS) - \frac{\ba\Th\matS(\matI - \matS)\ba}{\ba\Th(\matI - \matS)\ba},\\
&= \tnr{Tr}(\matS) - \ba\Th\matS\hat\bw,\\
&= M - \alpha\tnr{Tr}[(\matR_{\bx} + \alpha\matI)^{-1}] - \ba\Th\matR_{\bx}(\matR_{\bx} + \alpha\matI)^{-1}\hat\bw.
\end{align}

\ifdefined\ARXIV
\input{\CFilesBib/output.bbl.my}
\else
\bibliography{\CFilesBib/references_rank,\CFilesBib/IEEEabrv,\CFilesBib/references_all,\CFilesBib/wiener}
\bibliographystyle{\CFilesBib/IEEEtran}
\fi

\end{document}